\documentclass[12pt]{article}
\usepackage{epsfig}
\usepackage{amssymb}
\usepackage{amsmath}
\usepackage{amsfonts}
\usepackage{graphicx}
\usepackage{mathrsfs}
\usepackage[dvips]{color}
\usepackage{multirow}

\newcommand{\bsigma}{\boldsymbol{\sigma}}

\newcommand{\bnabla}{\boldsymbol{\nabla}}

\newcommand{\R}{\mathbb{R}}
\newcommand{\C}{\mathbb{C}}

\newcommand{\fb}{\mathfrak{b}}
\newcommand{\fc}{\mathfrak{c}}

\newcommand{\ff}{\mathfrak{f}}

\newcommand{\fz}{\mathfrak{z}}

\newcommand{\fM}{\mathfrak{M}}

\newcommand{\bk}{\mathbf{k}}

\newcommand{\bfr}{\mathbf{r}}
\newcommand{\bp}{\mathbf{p}}
\newcommand{\bx}{\mathbf{x}}

\newcommand{\bI}{\mathbf{I}}

\newcommand{\bM}{\mathbf{M}}

\newcommand{\cH}{\mathcal{H}}

\newcommand{\cF}{\mathcal{F}}

\newcommand{\cK}{\mathcal{K}}

\newcommand{\cM}{\mathcal{M}}

\newcommand{\cU}{\mathcal{U}}

\newcommand{\be}{\begin{equation}}
\newcommand{\ee}{\end{equation}}
\newcommand{\bea}{\begin{eqnarray}}
\newcommand{\eea}{\end{eqnarray}}
\newcommand{\nn}{\nonumber}
\newcommand{\kt}{\rangle}
\newcommand{\br}{\langle}

\newcommand{\ed}{\end{document}}

\newcommand{\bi}{\begin{itemize}}
\newcommand{\ei}{\end{itemize}}

\newcommand{\bce}{\begin{center}}
\newcommand{\ece}{\end{center}}
\newcommand{\sA}{\mathscr{A}}
\newcommand{\sB}{\mathscr{B}}

\newcommand{\sF}{\mathscr{F}}

\newcommand{\bPi}{{\boldsymbol{\Pi}}}
\newcommand{\bcK}{{\boldsymbol{\cK}}}
\newcommand{\bcM}{{\boldsymbol{\cM}}}

\newcommand{\bfM}{{\boldsymbol{\fM}}}
\newcommand{\bcH}{{\boldsymbol{\cH}}}
\newcommand{\bcU}{{\boldsymbol{\cU}}}

\newcommand{\bzero}{{\boldsymbol{0}}}

\newcommand{\for}{{\mbox{\rm for}}}

\newcommand{\bal}{\begin{align}}
\newcommand{\eal}{\end{align}}

\begin{document}



\title{Singularity-free treatment of delta-function point scatterers in two dimensions and its conceptual implications}

\author{Farhang Loran\thanks{E-mail address: loran@iut.ac.ir}~ and
Ali~Mostafazadeh\thanks{E-mail address:
amostafazadeh@ku.edu.tr}\\[6pt]
$^{*}$Department of Physics, Isfahan University of Technology, \\ Isfahan 84156-83111, Iran\\[6pt]
$^\dagger$Departments of Mathematics and Physics, Ko\c{c}
University,\\  34450 Sar{\i}yer, Istanbul, Turkey}

\date{ }
\maketitle

\begin{abstract}
In two dimensions, the standard treatment of the scattering problem for a delta-function potential, $v(\mathbf{r})=\mathfrak{z}\,\delta(\mathbf{r})$, leads to a logarithmic singularity which is subsequently removed by a renormalization of the coupling constant $\mathfrak{z}$. Recently, we have developed a dynamical formulation of stationary scattering (DFSS) which offers a singularity-free treatment of this potential. We elucidate the basic mechanism responsible for the implicit regularization property of DFSS that makes it avoid the logarithmic singularity one encounters in the standard approach to this problem. We provide an alternative interpretation of this singularity showing that it arises, because the standard treatment of the problem takes into account contributions to the scattered wave whose momentum is parallel to the detectors' screen. The renormalization schemes used for removing this singularity has the effect of subtracting these unphysical contributions, while DFSS has a built-in mechanics that achieves this goal.
  


\end{abstract}

\section{Introduction}

{The study of point interactions modeled using a delta-function potential has a long history. It began immediately after the formulation of quantum mechanics \cite{KP} and has continued to the present day. There are two main reasons for the interest in delta-function potentials; they have concrete applications in different areas of physics \cite{KP,Fermi,Foldy-1945,Lieb-1963a, Lieb-1963b,Antonie-1994,VCL,cervero,batchelor,josa-2020}, and at the same time admit exact analytic treatments which makes them an effective tool for teaching concepts and methods of quantum mechanics. In this respect there is a major difference between delta-function potentials in one dimension \cite{flugge} and those in two and three dimensions \cite{Albaverio}.}

The standard solution of the scattering problem for a delta-function potential in two dimensions is plagued with the emergence of singularities that are reminiscent of those encountered in quantum field theories. These singularities can be easily regularized and removed by adopting a coupling constant renormalization \cite{thorn,jackiw,mead,manuel,Adhikari1,Adhikari2,Mitra,Rajeev-1999,Nyeo,Camblong,teo-2006,teo-2010,teo-2013,ferkous,ap-2019}. This has made delta-function potentials in two dimensions into an ideal pedagogical tool for teaching the basic idea and methods of renormalization theory \cite{mead,Mitra,Nyeo}. Recently, we have proposed a solution of the scattering problem for these potentials that avoids the singularities of their standard treatment and yields the correct formula for their scattering amplitude  \cite{pra-2016,pra-2021}. This follows as an application of a dynamical formulation of stationary scattering (DFSS) where the solution of a scattering problem is related to the evolution operator for an effective non-unitary quantum system \cite{pra-2016,pra-2021}.
The fact that the application of DFSS to delta-function potentials in two (and three) dimensions does not encounter any singularities reveals an implicit regularization property of DFSS. The purpose of the present article is to elucidate the basic mechanism responsible for this property.

The first step toward revealing the origin of the mysterious implicit regularization property of DFSS is a careful examination of the standard treatment of the scattering problem for the delta-function potential in two dimensions. In the remainder of this section we offer a self-contained review of the latter.

Consider the time-independent Schr\"odinger equation
	\be
	-\nabla^2\psi(\bfr)+v(\bfr)\psi(\bfr)=k^2\psi(\bfr),
	\label{sch-eq}
	\ee
for the delta-function potential,
	\be
	v(\bfr)=\fz\,\delta(\bfr),
	\label{delta}
	\ee
where $\bfr$ is the position vector {in two dimensions}, $k$ is the wavenumber, and $\fz$ is a real or complex coupling constant. Let $|\bzero\kt$ denote the position ket $|\bfr\kt$ for $\bfr=\bzero$. Then the delta-function potential (\ref{delta}) is the position representation of the operator, $\hat v=\fz|\bzero\kt\br \bzero|$. Substituting this relation in the Lippmann-Schwinger equation, $|\psi\kt=|\bk\kt+\hat G\hat v |\psi\kt$, we have
	\be
	\br\bfr|\psi\kt=\br\bfr|\bk\kt+\fz\,G(\bfr)\br\bzero|\psi\kt,
	\label{LS-eq}
	\ee
where $\bk$ is the incident wave vector, $\hat G:=\lim_{\epsilon\to 0^+}\left(k^2-\hat\bp^2+i\epsilon\right)^{-1}$, $\hat\bp$ is the standard momentum operator, 
	\be
	G(\bfr):=\br\bfr|\hat G|\bzero\kt=\lim_{\epsilon\to 0^+}\int_{\R^2}
	\frac{d^2\bp}{(2\pi)^2}\:\frac{e^{i\bfr\cdot\bp}}{k^2-\bp^2+i\epsilon}=
	-\frac{i}{4}\,H^{(1)}_0(kr),
	\label{Green}
	\ee
$H^{(1)}_0(x)$ is the zero-order Hankel function of the first kind, and $r:=|\bfr|$.

Eq.~(\ref{LS-eq}) together with the asymptotic expression for the $H^{(1)}_0(x)$ and the fact that $\br\bfr|\bk\kt=e^{i\bk\cdot\bfr}/2\pi$ imply
	\be
	\br\bfr|\psi\kt\to\frac{e^{i\bk\cdot\bfr}}{2\pi}
	-\fz\br\bzero|\psi\kt\, \sqrt{\frac{i}{8\pi k r}}\,e^{ikr} 
	~~~\for~~~r\to\infty.
	\label{asymp}
	\ee
If we identify the scattering amplitude $\ff(\bk',\bk)$ through the following asymptotic expression for the solutions of the Lippmann-Schwinger equation \cite{adhikari},
	\be
	\br\bfr|\psi\kt\to
	\frac{1}{2\pi}\left[e^{i\bk\cdot\bfr}+\sqrt{\frac{i}{kr}}\,e^{ikr} \ff(\bk',\bk) \right]
	~~~\for~~~r\to\infty,
	\label{asymp-ff}
	\ee
where $\bk':=k\bfr/r$ is the scattered wave vector, we can use (\ref{asymp}) to deduce,
	\be
	\ff(\bk',\bk)=-\sqrt{\frac{\pi}{2}}\,\fz\br\bzero|\psi\kt.
	\label{ff=1}
	\ee
This equation reduces the determination of the scattering amplitude to the calculation of $\br\bzero|\psi\kt$. We can do this simply by setting $\bfr=\bzero$ in (\ref{LS-eq}), which gives $\fz\br\bzero|\psi\kt=\{2\pi[\fz^{-1}-G(\bzero)]\}^{-1}$. Substituting this equation in (\ref{ff=1}), we find
	\be
	\ff(\bk',\bk)=-\sqrt{\frac{1}{8\pi}}\,\frac{1}{\fz^{-1}-G(\bzero)}.
	\label{ff=2}
	\ee
The difficulty with this formula is that $G(\bzero)$ which is proportional to $H_0^{(1)}(0)$ diverges logarithmically. More specifically, we have
	\be
	H_0^{(1)}(x)=\frac{2i}{\pi}\left(\ln\frac{x}{2}+\gamma\right)+1+O(x^2),
	\label{H-asymp}
	\ee
where $\gamma$ is the Euler number, and $O(x^m)$ stands for terms of order $m$ and higher in powers of $x$.

To remove the divergent term entering (\ref{ff=2}), we regularize $G(\bfr)$ and use it to absorb the singularity of $G(\bzero)$ in $\fz^{-1}$. This requires interpreting $\fz$ as a bare coupling constant which has no physical significance. To make this explicit we use $\mathring\fz$ for $\fz$, so that (\ref{ff=2}) reads
	\be
	\ff(\bk',\bk)=-\sqrt{\frac{1}{8\pi}}\,\frac{1}{\mathring\fz^{-1}-G(\bzero)}.
	\label{ff=2-bare}
	\ee

We can regularize $G(\bzero)$ by expressing the integral in (\ref{Green}) in polar coordinate $(p,\varphi)$ in the momentum space, evaluate the angular integral, and put a cut-off on the radial coordinate $p$. In this way, we find 
	\be
	G(\bfr)\longrightarrow G_\Lambda(\bfr):=\lim_{\epsilon\to 0^+}\int_0^\Lambda \frac{dp}{2\pi}\:
	\frac{p\,J_0(pr)}{k^2-p^2+i\epsilon},
	\label{G-Lambda}
	\ee
so that $G(\bfr)=\lim_{\Lambda\to\infty}G_\Lambda(\bfr)$ for $\bfr\neq\bzero$. It is also easy to show that
	\bea
	G_\Lambda(\bzero)&=&-\frac{1}{4\pi}\ln\left(\frac{\Lambda^2}{k^2}-1\right)-\frac{i}{4}=
	-\frac{1}{2\pi}\ln\left(\frac{\Lambda}{k}\right)-\frac{i}{4}+O\big((k/\Lambda)^2\big).
	\label{G-mu-zero}
	\eea
According to (\ref{Green}), (\ref{H-asymp}), and (\ref{G-mu-zero}), for every positive real number $\alpha$, 
	\bea
	\lim_{r\to 0}\Big[G(\bx)-G_{_{\!\mbox{\footnotesize$\frac{\alpha}{r}$}}}\!(\bzero)\Big]=\frac{\gamma+\ln(\alpha/2)}{2\pi}.
	\label{G=G+}
	\eea
Therefore, we can identify $G(\bx)$ with $G_{_{\!\mbox{\footnotesize$\frac{\alpha}{r}$}}}\!(\bzero)+[\gamma+\ln(\alpha/2)]/2\pi$ whenever $r\ll\alpha/k$. 

Next, we introduce an arbitrary momentum scale $\mu$, set $\alpha:=2 e^{-\gamma}\mu/k$, and introduce the renormalized coupling constant,
	\be
	\tilde\fz:=\left(\mathring\fz^{-1}+\frac{1}{2\pi}\ln\frac{\Lambda}{\mu}\right)^{-1}=
	\left(\mathring\fz^{-1}+\frac{1}{2\pi}\ln\frac{\Lambda}{k}-\frac{\gamma+\ln(\alpha/2)}{2\pi}\right)^{-1}.
	\label{zn-renorm-1}
	\ee
Supposing that $\mathring\fz$ depends on $\Lambda$ in such a way that $\tilde\fz$ is $\Lambda$-independent, we can use (\ref{G-mu-zero}) -- (\ref{zn-renorm-1}) to show that, for $r=\alpha/\Lambda\to 0$,
	\be
	\mathring\fz^{-1}-G(\bx)~
	\longrightarrow~  \mathring\fz^{-1}-G_{\Lambda}(\bzero)-\frac{\gamma+\ln(\alpha/2)}{2\pi}=
	\tilde\fz^{-1}+\frac{i}{4}.\nn
	\ee
Making use of this observation and Eq.~(\ref{ff=2-bare}), we arrive at the following expression for the scattering amplitude of the delta-function potential (\ref{delta}).
	\be
	\ff(\bk',\bk)=-\sqrt{\frac{1}{8\pi}}\,\frac{1}{\tilde\fz^{-1}+\frac{i}{4}}.
	\label{ff-renorm}
	\ee
Notice that the value of the renormalized coupling constant $\tilde\fz$ is to be determined using experimental data. In principle, $\tilde\fz$ may depend on $\bk$. This suggests that the only physical prediction of (\ref{ff-renorm}) is the isotropic nature of the scattering amplitude, i.e., the fact that it does not dependent on $\hat\bk'=\bfr/r$.

\section{Dynamical formulation of stationary scattering in 2D}

\begin{figure}
        \begin{center}
        \includegraphics[scale=.25]{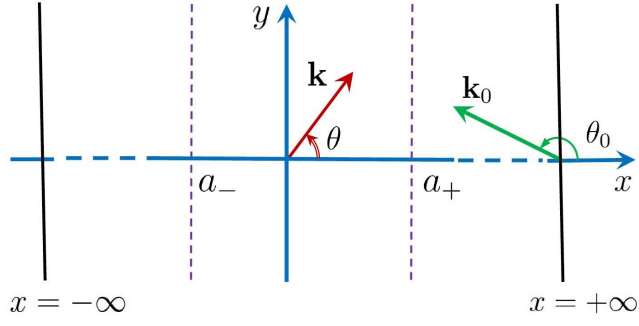}
        \caption{Schematic view of the scattering setup for a right-incident wave. 
        The solid black vertical lines represent spatial infinities where 
        the detectors are located. $\bk_0$ and $\bk$ are respectively the
        incident and scattered wave vectors. The incidence
        angle $\theta_0$ takes values in the interval $(\frac{\pi}{2},\frac{3\pi}{2})$. The
        support of the potential lies between the (dashed purple) lines $x=a_-$
        and $x=a_+$.}
        \label{fig1}
        \end{center}
        \end{figure}

DFSS has been developed in an attempt towards a generalization of the notion of transfer matrix to two and three dimensions \cite{pra-2016,pra-2021} that unlike the earlier generalizations \cite{pendry-1984,pendry-1990a,pendry-1990b,pendry-1994} does not involve a discretization of the transverse configuration or momentum space variables. It relies on the choice of a scattering axis which makes right angles with the lines at spatial infinity on which the source of the incident wave and the detectors of the scattering waves are located. We identify the scattering axis with the $x$-axis of our coordinate system and suppose that source of the incident wave and detectors reside on the lines $x=\pm\infty$. If the source is at $x=-\infty$ (respectively $x=+\infty$), we use the qualification ``left'' and ``right'' for the incident wave. Figure~\ref{fig1} shows a schematic view of the scattering setup for a right-incident wave.

Consider a potential $v:\R^2\to\C$ that vanishes in the region bounded by a pair of lines that are parallel to the $y$-axis, i.e., there are 
real numbers $a_\pm$ such that $a_-<a_+$ and
	\[v(x,y)=0~~~\for~~~x\notin [a_-,a_+].\]
It is easy to see that every bounded solution of the Schr\"odinger equation~(\ref{sch-eq}) for this potential satisfies 
    \be
    \psi(x,y)=\left\{\begin{array}{ccc}
    \displaystyle\int_{-\infty}^\infty\frac{dp}{4\pi^2\varpi(p)}\:
    \left[A_-(p)\,e^{i\varpi(p)x}+\sB_-(p)e^{-i\varpi(p)x}\right]e^{ipy}&\for&x\leq a_-,\\[12pt]
    \displaystyle\int_{-\infty}^\infty\frac{dp}{4\pi^2\varpi(p)}\:
    \left[\sA_+(p)\,e^{i\varpi(p)x}+B_+(p)e^{-i\varpi(p)x}\right]e^{ipy}&\for&x\geq a_+,\end{array}\right.
    \label{outside}
    \ee
where
    \begin{align}
    &\varpi(p):=\left\{\begin{array}{ccc}
    \sqrt{k^2-p^2}&\for&|p|<k,\\
    i\sqrt{p^2-k^2}&\for&|p|\geq k,
    \end{array}\right.
    \label{varpi-def}
    \end{align}
and $A_-,\sB_-,\sA_+$, and $B_+$ are complex-valued (generalized) functions\footnote{Ref.~\cite{pra-2021} uses $\breve A_-,\breve\sB_-,\breve\sA_+$, and $\breve B_+$ for what we call $A_-,\sB_-,\sA_+$, and $B_+$, respectively. Furthermore, the normalization of the wave function (\ref{outside}) adopted in Ref.~\cite{pra-2021} differs from the present article's by a factor of $2\pi$.} such that
    \begin{align}
    &A_-(p)=B_+(p)=0~~\for~~|p|\geq k.
    \label{A-B+}
    \end{align}
    
According to (\ref{outside}) -- (\ref{A-B+}), the coefficient functions $\sB_-$ and $\sA_+$ contain the information about the evanescent part of $\psi$ which decay exponentially as $x\to\pm\infty$. These are determined by $\sB_-(p)$ and $\sA_+(p)$ with $|p|\geq k$. The oscillating parts of $\psi$, which reach spatial infinity, are given by $A_-(p)$, $B_+(p)$, and $\sB_-(p)$ and $\sA_+(p)$ for $|p|<k$. We denote the latter by $B_-(p)$ and $A_+(p)$, respectively. In order to make the definitions of $B_-$ and $A_+$ more precise, we use $\sF$ to denote the set of complex-valued (generalized) functions possessing a Fourier transform, and let $\sF_k$ be the subset of $\sF$ given by
	\[\sF_k:=\{\phi\in\sF_k~|~\phi(p)=0~\for~|p|\geq k\},\]
and $\widehat\Pi_k:\sF\to\sF$ be the projection operator defined by
	\be
	(\widehat\Pi_k\phi)(p):=\left\{\begin{array}{ccc}
	\phi(p) & \for & |p|<k,\\
	0 & \for & |p|\geq k.\end{array}\right.
	\label{project}
	\ee
We can now define $B_-$ and $A_+$ via
	\begin{align}
    	&A_+:=\widehat\Pi_k\sA_+, &&B_-:=\widehat\Pi_k\sB_-.
    	\label{AB-proj-1}
    	\end{align}
In view of (\ref{outside}) -- (\ref{AB-proj-1}),
    \be
    \psi(x,y)\to\int_{-k}^k\frac{dp}{4\pi^2\varpi(p)}\left[A_\pm(p)\,e^{i\varpi(p)x}+
     B_\pm(p)\,e^{-i\varpi(p)x}\right]e^{ipy}~~\for~~x\to\pm\infty.
    \label{asym-2}
    \ee
Note also that $A_\pm,B_\pm\in\sF_k$.

By analogy to one dimension \cite{tjp-2020}, we identify the transfer matrix of $v$ with a $2\times 2$ matrix $\widehat{ \bM}$ with operator entries that satisfies \cite{pra-2021}
    \begin{align}
    &\widehat{\bM}\left[\begin{array}{c}
    A_-\\
    B_-\end{array}\right]=
    \left[\begin{array}{c}
    A_+\\
    B_+\end{array}\right].
    \label{M-def}
    \end{align}
It is important to realize that $\widehat{ \bM}$ is not a numerical matrix; it is a linear operator acting in the infinite-dimensional function space of two-component wave functions,
    \[\sF_k^{2\times 1}:=\left.\left\{\left[\begin{array}{c}
    \phi_+\\
    \phi_-\end{array}\right]~\right|~\phi_\pm\in\sF_k~\right\}.\]
    
A remarkable property of the transfer matrix $\widehat{\bM}$ is that it contains the information about the scattering properties of the potential. To describe this, we confine our attention to right-incident waves (whose source is located at $x=+\infty$). If we denote the $y$-component of the incident wave vector $\bk$ by $p_0$, its $x$-component is given by $-\varpi(p_0)$, and $\bk\cdot\bfr=-\varpi(p_0)x+p_0y$. Furthermore, for such a right-incident wave, $A_-$ and $B_+$ take the form,
	\begin{align}
    	&A_-(p)=0,
    	\quad\quad\quad
    	B_+(p)=2\pi\varpi(p_0)\,\delta(p-p_0),
     	\label{ABs}
     	\end{align}
and consequently (\ref{outside}) and (\ref{asym-2}) become
	\begin{align}
	\psi(x,y)=
	\left\{\begin{array}{ccc}
	\displaystyle
	\int_{-\infty}^\infty \frac{dp}{4\pi^2\varpi(p)}\,\sB_-(p)\,
	e^{-i\varpi(p)x}e^{ipy}&\for&x\leq a_-,\\[12pt]
	\displaystyle\frac{e^{i\bk\cdot\bfr}}{2\pi}+
	\displaystyle
	\int_{-\infty}^\infty
	\frac{dp}{4\pi^2\varpi(p)}\: \sA_+(p)\,e^{i\varpi(p)x}e^{ipy}&\for&x\geq a_+,
	\end{array}\right.
    	\label{out-side2}
	\end{align}
and
	\begin{align}
	\psi(x,y)\to
	\left\{\begin{array}{ccc}
	\displaystyle
	\int_{-k}^k \frac{dp}{4\pi^2\varpi(p)}\,B_-(p)\,
	e^{-i\varpi(p)x}e^{ipy}&\for&x\to-\infty,\\[12pt]
	\displaystyle\frac{e^{i\bk\cdot\bfr}}{2\pi}+
	\displaystyle
	\int_{-k}^k\frac{dp}{4\pi^2\varpi(p)}\: A_+(p)\,e^{i\varpi(p)x}e^{ipy}&\for&x\to+\infty.
	\end{array}\right.
    	\label{asym-3}
	\end{align}
		
Let $(k,\theta_0)$ and $(k,\theta)$ denote the polar coordinates of incident and scattering wave vectors, $\bk$ and $\bk'$, respectively. The scattering amplitude $\ff(\bk',\bk)$ is clearly a function of $k$, $\theta_0$, and $\theta$. In the following we keep the dependence of $\ff(\bk',\bk)$ on $k$ and $\theta_0$ implicit, and use $\ff(\theta)$ to denote $\ff(\bk',\bk)$ for brevity, i.e.,
	\[\ff(\theta):=\ff(\bk',\bk).\]
Notice also that because we consider right-incident waves, whose source is at $x=+\infty$, $\theta_0\in(\frac{\pi}{2},\frac{3\pi}{2})$. 	
	
Comparing the asymptotic expressions (\ref{asymp-ff}) and (\ref{asym-3}) and using a result derived in \cite[Appendix~A]{pra-2016}, we can show that 
        \begin{align}
        &\ff(\theta)=-\frac{i}{\sqrt{2\pi}}\left\{
        \begin{array}{ccc}
        A_+(k\sin\theta) &\for & \theta\in(-\frac{\pi}{2},\frac{\pi}{2}),\\[3pt]            
        B_-(k\sin\theta)-2\pi\delta(\theta-\theta_0)  &\for & \theta\in(\frac{\pi}{2},\frac{3\pi}{2}).\end{array}\right.
    \label{f-R}
        \end{align}
Next, we substitute (\ref{ABs}) in (\ref{M-def}). This gives
	  \begin{align}
     &\widehat{ M}_{22} B_-=2\pi\varpi(p_0)\,\delta_{p_0},
     && A_+=\widehat{ M}_{12} B_-,
        \label{e2}
     \end{align}
where $\delta_{p_0}$ denotes the delta-function centered at $p_0$, i.e., $\delta_{p_0}(p):=\delta(p-p_0)$. Eqs.~(\ref{f-R}) and (\ref{e2}) reduce the determination of the scattering amplitude to the calculation of the transfer matrix and the solution of the first equation in (\ref{e2}). 

We use the term DFSS for this approach to stationary scattering, because we can express the transfer matrix $\widehat\bM$ in terms of the evolution operator for an effective non-unitary quantum system. A proper description of this feature of $\widehat\bM$ requires the introduction of a linear operator, which we call the ``auxiliary transfer matrix'' and denote by $\widehat\bfM$, \cite{pra-2021}. This is an operator that acts in the space of two-component wave functions,
	\[\sF^{2\times 1}:=\left.\left\{\left[\begin{array}{c}
    	\xi_+\\
    	\xi_-\end{array}\right]~\right|~\xi_\pm\in\sF~\right\},\]
and satisfies
	\begin{align}
    &\widehat{\bfM}\left[\begin{array}{c}
	A_-\\
	\sB_-\end{array}\right]=
	\left[\begin{array}{c}
	\sA_+\\
	B_+\end{array}\right].
	\label{M-def-aux}
	\end{align}
The auxiliary transfer matrix has the following important properties \cite{pra-2021}. 
\begin{enumerate}
\item In view of (\ref{ABs}) and (\ref{M-def-aux}), the coefficient functions $\sB_-$ and $\sA_+$ for a right-incident wave satisfy
	 \begin{align}
	&\widehat{\fM}_{22}\sB_-=2\pi\varpi(p_0)\,\delta_{p_0},
	&& \sA_+=\widehat{\fM}_{12}\sB_-.
        	\label{e2-axiliary}
        	\end{align}
If we can solve the first of these equations for $\sB_-$ and substitute the result in the second, we obtain $\sB_-$ and $\sA_+$. For $|p|<k$, (\ref{AB-proj-1}) implies that $B_-(p)=\sB_-(p)$ and $A_+(p)=\sA_+(p)$. Using these relations in (\ref{f-R}), we can express the scattering amplitude for the right-incident waves as
	\be
	\ff(\theta)=-\frac{i}{\sqrt{2\pi}}\left\{
        \begin{array}{ccc}
        \sA_+(k\sin\theta) &\for & \theta\in(-\frac{\pi}{2},\frac{\pi}{2}),\\[3pt]            
        \sB_-(k\sin\theta)-2\pi\delta(\theta-\theta_0)  &\for & \theta\in(\frac{\pi}{2},\frac{3\pi}{2}).\end{array}\right.
    \label{f-R-aux}
	\ee
	
\item The auxiliary transfer matrix admits an expression in terms of the evolution operator $\widehat\bcU(x,x_0)$ for the Hamiltonian operator,		
	\be
   	\widehat\bcH(x):=\frac{1}{2}\,e^{-i\widehat\varpi x\bsigma_3}
         v(x,\hat y)\,\widehat\varpi ^{-1}\bcK\, e^{i\widehat\varpi x\bsigma_3},
         \label{bcH-def}
        \ee
where $x$ plays the role of time, $\widehat\varpi:=\varpi (\hat p)=\int_{-\infty}^\infty dp\:\varpi(p)|p\kt\br p|$,
$\hat y$ and $\hat p$ are respectively the $y$-component of the standard position and momentum operators, i.e.,
$(\hat y\phi)(p)=i\partial_p\phi(p)$ and $(\hat p \phi)(p):=p \phi(p)$, 
	\be
	\bcK:=\left[\begin{array}{cc}
    1 & 1 \\
    -1 & -1\end{array}\right]=\bsigma_3+i\bsigma_2,
    \label{bcK-def}
    \ee
and $\bsigma_j$, with $j\in\{1,2,3\}$, denote the Pauli matrices. We can view $v(x,\hat y)$ as the operator acting in $\sF$ according to
    \be
    \big(v(x,\hat y)\phi\big)(p):=\frac{1}{2\pi}\int_{-\infty}^\infty dq\:
    \tilde v(x,p-q)\phi(q),
    \label{v-op1}
    \ee
where a tilde over a function of $(x,y)$ stands for its Fourier transform with respect to $y$, i.e., $\tilde f(x,p):=\int_{-\infty}^\infty dy\: e^{-ipy}f(x,y)$.	The evolution operator $\widehat\bcU(x,x_0)$ for the Hamiltonian~(\ref{bcH-def}) gives the auxiliary transfer matrix according to $\widehat\bfM=\widehat\bcU(a_+,a_-)$. In particular, employing the Dyson series expansion of $\bcU(x,x_0)$ and noting that $\widehat{\bcH}(x)=\bzero$ for $x\notin[a_-,a_+]$, we have		
	\be
	\widehat\bfM=\widehat\bI+\sum_{n=1}^\infty (-i)^n
         \int_{-\infty}^{\infty} \!\!dx_n\int_{-\infty}^{x_n} \!\!dx_{n-1}
         \cdots\int_{-\infty}^{x_2} \!\!dx_1\,
         \widehat{\bcH}(x_n)\widehat{\bcH}(x_{n-1})\cdots\widehat{\bcH}(x_1).
         \label{bcM-Dyson}
	\ee
As shown in Ref.~\cite{pra-2021}, we can use $\widehat\bfM=\widehat\bcU(a_+,a_-)$ to establish the composition property of $\widehat\bfM$ which is reminiscent of the well-known composition property of the transfer matrix of scattering theory in one dimension \cite{tjp-2020}. 

\item The auxiliary transfer matrix $\widehat\bcM$ is related to the (fundamental) transfer matrix $\widehat\bM$ according to
	\be
    	\widehat\bM=\widehat\bPi_k\,\widehat\bfM\,\widehat\bPi_k,
    	\label{M-M}
    	\ee
where $\widehat\bPi_k$ is the projection operator defined on $\sF^{2\times 1}$ according to
    	\be
    	\widehat\bPi_k\left[\begin{array}{c}\xi_+\\ \xi_-\end{array}\right]:=
    	\left[\begin{array}{c}\widehat\Pi_k\xi_+\\ \widehat\Pi_k\xi_-\end{array}\right]~~~
    	\for~~~\left[\begin{array}{c}\xi_+\\ \xi_-\end{array}\right]\in\sF^{2\times 1},
    	\label{Pik-def}
    	\ee
and $\widehat\Pi_k$ is defined by (\ref{project}).
	
\end{enumerate}

\section{Application to delta-function potential}

For the delta-function potential~(\ref{delta}), {which we can express as $v(x,y)=\fz\,\delta(x)\delta(y)$}, the Hamiltonian operator (\ref{bcH-def}) takes the simple form,
	\be
   	\widehat\bcH(x):=\frac{1}{2}\,v(x,\hat y)\widehat\varpi ^{-1}\,\bcK,
         \label{bcH-def-delta}
        \ee
{where we have used the $\delta(x)$ in the expression for $v(x,\hat y)$ to replace the $e^{\pm i \widehat\varpi x\bsigma_3}$ in (\ref{bcH-def}) with the identity operator.} Because $\bcK^2=\bzero$, (\ref{bcH-def-delta}) implies that $\widehat\bcH(x_1)\widehat\bcH(x_2)=\bzero$ for all $x_1,x_2\in\R$. Therefore the Dyson series on the right-hand side of (\ref{bcM-Dyson}) truncates, and we find
	\be
	\widehat\bfM=\bI-\frac{i}{2}\int_{-\infty}^{\infty}
	dx\:v(x,\hat y)\widehat\varpi ^{-1}\,\bcK.
	\label{cfM-deta}
	\ee

Next, we determine the explicit form of the operator $\tilde v(x,p)$ for the delta-function potential~(\ref{delta}). Evaluating the Fourier transform of this potential and using it in (\ref{v-op1}), we obtain, $\tilde v(x,p)=\fz\,\delta(x)$, and
	\begin{align}
	\big(v(x,\hat y)\phi\big)(p)=\frac{\fz\,\delta(x)}{2\pi}\int_{-\infty}^\infty dq\:\phi(q).
	\label{v-op2}
	\end{align}
Substituting this equation in (\ref{cfM-deta}) and employing (\ref{bcK-def}), we can calculate the entries of $\widehat\bfM$. In particular, this gives
	\begin{align}
	&\big(\fM_{12}\phi\big)(p)= -\frac{i\fz}{4\pi}\int_{-\infty}^\infty dq\:\frac{\phi(q)}{\varpi(q)},
	&&\big(\fM_{22}\phi\big)(p)= \phi(p)+\frac{i\fz}{4\pi}\int_{-\infty}^\infty dq\:\frac{\phi(q)}{\varpi(q)}.
	\label{axi-sol}
	\end{align}
Plugging these equations in (\ref{e2-axiliary}), we are led to 
	\begin{align}
	&\sA_+(p)=\fc,
	\label{sAp=}\\
	&\sB_-(p)=2\pi\varpi(p_0)\delta(p-p_0)+\fc.
	\label{sBm=}
	\end{align}
where
	\be
	\fc:=-\frac{i\fz}{4\pi} \int_{-\infty}^\infty
	dq\:\frac{\sB_-(q)}{\varpi(q)}. 
	\label{c-def}
	\ee
In view of (\ref{out-side2}), (\ref{f-R-aux}), (\ref{sAp=}), and (\ref{sBm=}),
	\begin{align}
	\psi(x,y)=\frac{e^{i\bk\cdot\bfr}}{2\pi}+\frac{\fc}{4\pi^2}
	\int_{-\infty}^\infty \frac{dp}{\varpi(p)}\,e^{i\varpi(p)|x|}e^{ipy}~~~\for~~~r\neq 0,
	\label{psi-delta}
	\end{align}
and 
	\be
	\ff(\theta)=-\frac{i\fc}{\sqrt{2\pi}}
	\label{ff=alternative}.
	\ee
If we express the integrand in (\ref{Green}) in the Cartesian coordinates and perform the integral over the $x$-component of $\bp$, we find 
	\be
	\int_{-\infty}^\infty \frac{dp}{\varpi(p)}\,e^{i\varpi(p)|x|}e^{ipy}=\pi H_0^{(1)}(kr).
	\label{int-id1}
	\ee
This in turn allows us to write (\ref{psi-delta}) as
	\begin{align}
	\psi(x,y)=\frac{e^{i\bk\cdot\bfr}}{2\pi}+\frac{\fc}{4\pi}\,H_0^{(1)}(kr)
	~~~\for~~~r\neq 0,
	\label{psi-delta-2}
	\end{align}
which, in light of (\ref{Green}), coincides with the Lippmann-Schwinger equations (\ref{LS-eq}).
	
We can try to compute the value of the constant $\fc$ by substituting (\ref{sBm=}) in  (\ref{c-def}). The result is
	\be
	\fc=-\frac{i}{2}\left[\fz^{-1}+
	\frac{i}{4\pi}\int_{-\infty}^\infty\frac{dq}{\varpi(q)}\right]^{-1}.
	\label{c=}
	\ee
This calculation is however invalid, because the integral on the right-hand side of (\ref{c=}) blows up; in view of (\ref{int-id1}), it is equal $\pi\,H^{(1)}_0(0)$.
This is precisely the same singularity we encountered in Sec.~1. Therefore, the direct application of the auxiliary transform matrix for the solution of the scattering problem for the delta-function potential~(\ref{delta}) seems to be equivalent to the standard treatment of this problem in the sense that its proper implementation requires a regularization of a logarithmic singularity and a renormalization of the coupling constant. These turn (\ref{c=}) into 
	\be
	\fc= -\frac{i}{2\left(\tilde\fz^{-1}+\frac{i}{4}\right)},
	\label{c=123}
	\ee
which in view of (\ref{ff=alternative}) reproduces (\ref{ff-renorm}). 

If we substitute (\ref{c=123}) in (\ref{psi-delta-2}), we can use (\ref{H-asymp}) to show that	
	\[\psi(\bfr)=\frac{1}{4\pi^2\left(\tilde\fz^{-1}+\frac{i}{4}\right)}
	\Big[\ln\frac{kr}{2}+2\pi\tilde\fz^{-1}+\gamma+O(kr)^2\Big]~~~\for~~~kr\ll 1.\]
Therefore, $\psi(\bfr)$ has a logarithmic singularity at $r=0$ whose strength is determined by $\tilde\fz$. This confirms the link to the treatment of the delta-function potential (with a real coupling constant) that makes use of the theory of self-adjoint extensions \cite{jackiw}. 
	
Next, we compute the fundamental transfer matrix for the delta-function potential~(\ref{delta}). To do this we first use (\ref{M-M}) and (\ref{cfM-deta}) to express the fundamental transfer matrix in the form
	\be
	\widehat\bM=\widehat\bPi_k-\frac{i}{2}\widehat\bPi_k\int_{-\infty}^{\infty}
	dx\: v(x,\hat y)\widehat\varpi ^{-1}\,\bcK\,\widehat\bPi_k.
	\label{bM-deta}
	\ee
Then, in view of (\ref{project}), (\ref{bcK-def}), (\ref{Pik-def}), (\ref{v-op2}), (\ref{bM-deta}), we have
	\begin{align}
	&\big(\widehat M_{12}\phi\big)(p)=-\frac{i\fz}{4\pi} \int_{-k}^k dq\:\frac{\phi(q)}{\varpi(q)},
	&&\big(\widehat M_{22}\phi\big)(p)=\phi(p)+
	\frac{i\fz}{4\pi} \int_{-k}^k dq\:\frac{\phi(q)}{\varpi(q)}.
	\label{fund-sol}
	\end{align}
These relations allow us to write (\ref{e2}) as
	\begin{align}
	&A_+(p)=\fc',
	&&B_-(p)=2\pi\varpi(p_0)\delta(p-p_0)+\fc',
	\label{second}
	\end{align}
where
	\be
	\fc':=-\frac{i\fz}{4\pi}\int_{-k}^k\frac{B_-(q)}{\varpi(q)}=-\frac{i\fz}{4\pi}\int_{-k}^k\frac{B_-(q)}{\sqrt{k^2-q^2}}.
	\label{c-prime}
	\ee
Substituting the second of Eqs.~(\ref{second}) in the right-hand side of  relation (\ref{c-prime}), solving the result for $\fc'$, and using the identity, $\int_{-k}^k dq/\sqrt{k^2-q^2}=\pi$, we arrive at
	\be
	\fc'=-\frac{i}{2\left(\fz^{-1}+\frac{i}{4}\right)}.
	\label{c-prime=}
	\ee
Eqs.~(\ref{f-R}), (\ref{second}), and (\ref{c-prime=}) give the following expression for the scattering amplitude.
	\begin{align}
	\ff(\theta)=-\frac{1}{\sqrt{8\pi}}\,\frac{1}{\fz^{-1}+\frac{i}{4}}.
        \label{ff=dfss}
     	\end{align}
Remarkably, this formula coincides with (\ref{ff-renorm}), if we replace the coupling constant $\fz$ with the renormalized coupling constant $\tilde\fz$. This observation reveals a conceptual dilemma; if we solve the scattering problem using the auxiliary transfer matrix we arrive at (\ref{ff-renorm}), and we need to interpret $\fz$ as an unphysical bare coupling constant $\mathring\fz$ and renormalize it, but if we use the fundamental transfer matrix, we must treat $\fz$ as a physical parameter. These two interpretations are clearly incompatible! {This problem calls for a careful examination of the origin of the infinite quantity associated with the direct use of auxiliary transfer matrix and the reason why it does not emerge when we employ the fundamental transfer matrix.} 

{Before addressing this problem, which is necessary for the self-consistency of DFSS, we wish to elaborate on the compatibility of the physical outcomes of the DFSS and the standard treatment of the delta-function potential which we reviewed in Sec.~1. This again amounts to comparing Eqs.~(\ref{ff-renorm}) and (\ref{ff=dfss}), but this time the finite coupling constant $\fz$ of DFSS does not enter in the analysis of the standard treatment of this potential. The latter 
begins with replacing $\fz$ of the potential (\ref{delta}) with a bare coupling constant $\mathring\fz$, i.e., setting $v(\bfr)=\mathring\fz\,\delta(\bfr)$, and ends up with the expression (\ref{ff-renorm}) for the scattering amplitude which involves the renormalized coupling constant $\tilde\fz$. The derivation of (\ref{ff=dfss}) makes use of the formula (\ref{delta}) with $\fz$ having a finite value and yields  Eq.~(\ref{ff=dfss}) for the scattering amplitude. In principle, there is no relationship between the finite coupling constant $\fz$ of DFSS and the renormalized coupling constant $\tilde\fz$ of the standard approach, except that they appear in the expression for the scattering amplitude for the same point scatterer. To determine whether (\ref{ff-renorm}) or (\ref{ff=dfss}) conform with experiments, we should fix the numerical values of $\tilde\fz$ and $\fz$. This also requires experimental input. For example, we can fit the experimental value of the scattering amplitude at a specific angle $\theta_\star$ to (\ref{ff-renorm}) and (\ref{ff=dfss}) to 
determine $\tilde\fz$ and $\fz$. Clearly, this implies $\tilde\fz=\fz$ which renders (\ref{ff-renorm}) and (\ref{ff=dfss}) identical. This shows that both the standard approach and DFSS yield the same physical result. Their difference is the former leads to a singularity and requires the use of regularization and renormalization schemes to produce the physical result while the latter does not.}

\section{Origin of the implicit regularization property of DFSS}

The conceptual difficulty associated with the status of the coupling constant $\fz$ that we face in the preceding section has its origin in the way in which we remove the singular term $H^{(1)}_0(0)$ from the expression (\ref{ff=alternative}) for the scattering amplitude. In this section, we provide a completely different method of dealing with this singularity which does not require a renormalization of $\fz$. 

Let us write (\ref{sBm=}) in the form
	\be
	\sB_-(p)=2\pi\varpi(p_0)\delta(p-p_0)-\frac{i\fz}{4\pi} \int_{-\infty}^\infty
	dq\:\frac{\sB_-(q)}{\varpi(q)}.
	\label{sBm=2}
	\ee
This is an integral equation for $\sB_-$. We have previously denoted the last term on the right-hand side of (\ref{sBm=2}) by $\fc$ and offered a calculation of this quantity which yields (\ref{c=}). Substitution of this equation in (\ref{sBm=}) gives a solution of (\ref{sBm=2}) which involves the unwanted singularity. It turns out, however, that this equation does not have a unique solution. To see this, we let $\fb_\pm$ be a pair of complex parameters, and $\cF$ be a (generalized) function satisfying
	\be
	\cF(p)=2\pi \left[\delta(p-p_0)+\fb_+\delta(p-k)+\fb_-\delta(p+k)\right]-
	\frac{i\fz}{4\pi \varpi(p)} \int_{-\infty}^\infty
	dq\:\cF(q).
	\label{sBm=4}
	\ee
Because $\delta(p\mp k)\varpi(p)=\delta(p\mp k)\varpi(\pm k)=0$, the function $\sB_-$ given by,
	\be
	\sB_-(p):=\varpi(p)\cF(p),
	\label{new-soln}
	\ee
satisfies (\ref{sBm=2}). 

Next, we observe that according to (\ref{c-def}) and (\ref{new-soln}),
	\be
	\fc=-\frac{i\fz}{4\pi} \int_{-\infty}^\infty dq\:\cF(q).
	\label{ccc}
	\ee
We can use this relation to write (\ref{sBm=4}) in the form
	\be
	\cF(p)=2\pi \left[\delta(p-p_0)+\fb_+\delta(p-k)+\fb_-\delta(p+k)\right]+\frac{\fc}{\varpi(p)}.
	\label{sBm=5}	
	\ee
Substituting (\ref{sBm=5}) in the right-hand side of (\ref{ccc}), solving the resulting equation for $\fc$, and making use of (\ref{int-id1}), we find
	\be
	\fc=-\frac{i(1+\fb_-+\fb_+)}{2\left[\fz^{-1}+\frac{i}{4}H^{(1)}_0(0)\right]}.
	\label{ccc2}
	\ee
Eqs.~(\ref{new-soln}), (\ref{sBm=5}), and (\ref{ccc2}) identify the following two-parameter family of solutions of (\ref{sBm=2}).
	\be
	\sB_-(p)=2\pi\varpi(p_0)\delta(p-p_0)-
	\frac{i(1+\fb_-+\fb_+)}{2\left[\fz^{-1}+\frac{i}{4}H^{(1)}_0(0)\right]}.
	\ee
Furthermore, inserting (\ref{ccc2}) in (\ref{ff=alternative}), we obtain
	\be
	\ff(\theta)=-\frac{1}{\sqrt{8\pi}}\:
	\frac{1+\fb_-+\fb_+}{\fz^{-1}+\frac{i}{4}H^{(1)}_0(0)}.
	\label{ff-301}
	\ee	
This equation agrees with the outcome of the application of the fundamental matrix, namely (\ref{ff=dfss}) provided that
	\be
	\fz=\frac{4i}{1-\displaystyle\frac{H^{(1)}_0(0)-1}{\fb_-+\fb_+}}.
	\label{fz=302}
	\ee
Notice that this condition is equivalent to  
	\[\fc=\fc'=-\frac{i}{2\left(\fz^{-1}+\frac{i}{4}\right)}.\] 
	
According to (\ref{fz=302}), we can interpret $\fb_-+\fb_+$ as a bare parameter that is capable of absorbing the singularity $H^{(1)}_0(0)$ in such a way that the right-hand side of (\ref{fz=302}) remains finite and equals $\fz$. By construction, this will ensure that (\ref{ff-301}) coincides with (\ref{ff=dfss}). Therefore the coupling constant $\fz$ that enters the expression for the potential (\ref{delta}) is a physical parameter determining the scattering features of this potential. This argument shows that although the application of the auxiliary transfer matrix leads to a logarithmic singularity, we do not need to perform a renormalization of the coupling constant $\fz$ to subtract this singularity. Instead, we can require $\fb_-+\fb_+$ to absorb the singularity. 

Let $\Lambda:=r^{-1}$. Then, (\ref{H-asymp}) implies
	\[H_0^{(1)}(kr)-1=
	\frac{2i}{\pi}\ln\frac{\Lambda}{k}+O\big((k/\Lambda)^2\big)~~\for~~r\to 0.\]
In view of (\ref{fz=302}), this suggests the following renormalization of $\fb_-+\fb_+$.
	\be
	\fb_-+\fb_+\longrightarrow \tilde\fb:=\frac{i\pi(\fb_-+\fb_+)}{2\ln(\Lambda/k)}.
	\label{tilde-b}
	\ee
Enforcing this relation in (\ref{fz=302}), we find 
	\be
	\tilde\fb=\frac{\fz}{\fz-4i}.
	\nn
	\ee

In order to clarify the physical meaning of $\fb_\pm$, we note that according to (\ref{out-side2}), the solutions $\psi(\bfr)$ of the Schr\"odinger equation that correspond to right-incident waves are determined by $\sB_-(p)/\varpi(p)$ and $\sA_+(p)/\varpi(p)$. In view of (\ref{e2-axiliary}), (\ref{axi-sol}), (\ref{sBm=2}), and (\ref{sBm=5}), these satisfy
	\begin{align}
	&\frac{\sA_+(p)}{\varpi(p)}=\frac{\sB_-(p)}{\varpi(p)}-2\pi\delta(p-p_0)=
	2\pi \left[\fb_+\delta(p-k)+\fb_-\delta(p+k)\right]+\frac{\fc}{\varpi(p)}.
	\label{sA-sB=345}
	\end{align}
Eqs.~(\ref{out-side2}), (\ref{int-id1}), and (\ref{sA-sB=345}) imply
	\begin{align}
	&\psi(x,y)=\frac{e^{i\bk\cdot\bfr}}{2\pi}+\frac{\fc}{4\pi}H^{(1)}_0(kr)+\psi_0(y)
	~~~\for~~~r\neq 0,\nn
	\end{align}
where
	\be
	\psi_0(y):=\frac{1}{2\pi}(\fb_+ e^{iky}+\fb_- e^{-iky})=
	\frac{1}{2\pi}\left[(\fb_++\fb_-)\cos(ky)+i(\fb_+-\fb_-)\sin(ky)\right].
	\label{psi-0}
	\ee
The latter is a solution of the Schr\"odinger equation which can never affect the scattering properties of the potential.\footnote{This is because the probability current density, $-i(\psi^*\bnabla\psi-\psi\bnabla\psi^*)$, associated to (\ref{psi-0}) points along the $y$-axis. This makes it parallel to the detectors' screens that are placed on the lines $x=\pm\infty$.}
Therefore, as far as the scattering phenomenon is concerned, $\fb_\pm$ are unphysical parameters.  

Next, we recall that the application of the theory of self-adjoint extensions (for the cases where $\fz$ is real) shows that the
solutions of the Schr\"odinger equation for the delta-function potential~(\ref{delta}) has a logarithmic singularity at $\bfr=\bzero$,
\cite{jackiw}. Because for $\bfr\to \bzero$, $\psi_0(y)\to (\fb_++\fb_-)/2\pi$, this shows that $\fb_++\fb_-$ must be logarithmically divergent. This is consistent with (\ref{tilde-b}).

We do not encounter the singularity associated with $\psi_0$ while applying the fundamental transfer matrix, because by its very definition, namely (\ref{M-def}), it is only sensitive to the behavior of the coefficient functions $A_\pm$ and $B_\pm$ which describe the oscillating parts of the traveling wave solutions of the Schr\"odinger equation.

\section{Conclusions}

Renormalization theory has a special place among the major discoveries of the 20th century theoretical physics. Students usually begin learning it while studying statistical or quantum field theories. The bound-state and scattering problems for the delta-function potential in two and three dimensions provide valuable toy models where one can describe the basic idea and practical aspects of renormalization theory without having to deal with the typical complications arising in the study of field theories. Standard treatments of these potentials lead to singularities that could be removed via a renormalization of the coupling constant $\fz$ multiplying the delta function. This was recognized decades ago, and its various aspects and generalizations have been the subject of many research publications. The recent advent of the dynamical formulation of stationary scattering (DFSS) has however pointed at a different direction. It has revealed the possibility of a singularity-free treatment of the scattering problem for the delta-function potential in two and three dimensions \cite{pra-2016,pra-2021}. In this article we have outlined the details of this treatment and elucidated the basic mechanism that is responsible for its implicit regularization property.

A careful examination of the application of DFSS to delta-function potential in two dimensions shows that the logarithmic singularity that arises in its standard treatment is also present if one tries to compute the scattering amplitude using the auxiliary transfer matrix. But if one uses the fundamental transfer matrix for this purpose, this singularity does not enter into the calculations. In the former approach, we can again perform a coupling-constant renormalization which reproduces the result obtained using the approach based on the Lipmann-Schwinger equation. But this procedure leads to an inconsistency related to the status of the coupling constant $\fz$. The use of the auxiliary transfer matrix seems to identify $\fz$ with an unphysical parameter which can absorb the singularity, while the application of the fundamental transfer matrix treats $\fz$ as a physical parameter that determines the scattering amplitude of the potential. We have offered a resolution of this inconsistency by identifying a different bare parameter that can absorb the singularity arising in the application of the auxiliary transfer matrix. This is related to the $x$-independent solutions (\ref{psi-0}) of the Schr\"odinger equation which do not affect the scattering phenomenon. We have also shown that our findings are in agreement with the outcome of the theory of self-adjoint extensions. 

Because the solution of the scattering problem for the delta-function potential that relies on the auxiliary transfer matrix is equivalent to the standard treatment of this potential, our results suggest that the logarithmic singularity arising in these approaches stems from the inclusion of contributions to the scattered wave whose momentum is parallel to the detectors' screen. The renormalization schemes used for removing this singularity has the effect of subtracting these unphysical (undetectable) contributions. DFSS avoids the singularity, because it has a built-in mechanics for excluding these contributions.

Finally, we wish to note that the results pertaining the status of the coupling constant (as a physical/unphysical parameter), consistency of the singularity-free treatment of the delta-function potential offered by DFSS, and the interpretation of singular term arising in the standard treatment of point scatterers and the corresponding coupling-constant renormalization also apply in three dimensions.

 \section*{Acknowledgements}
This work has been supported by the Scientific and Technological Research Council of Turkey (T\"UB\.{I}TAK) in the framework of the project 120F061 and by Turkish Academy of Sciences (T\"UBA).

\ed